\documentclass{iopart}

\pdfoutput=1
\usepackage{graphicx} 
\usepackage{epsfig} 
 
\newcommand{\bma}{\begin{displaymath}} 
\newcommand{\ema}{\end{displaymath}}

\begin{document} 
 
\title[Metal cluster, quantum dot and quantum rings] 
{Electron correlation in metal clusters, quantum dots and quantum rings} 
 
\author{M Manninen$^1$, S M Reimann$^2$} 
 
\address{\sl $^1$NanoScience Center, Department of Physics, 
FIN-40014 University of Jyv\"askyl\"a, Finland} 
 
\address{\sl $^2$Mathematical Physics, Lund Institute of Technology, 
SE-22100 Lund, Sweden} 
  
 
\begin{abstract}  
This short review presents a few case studies of finite electron systems  
for which strong correlations play a dominant role. 
In simple metal clusters, the valence 
electrons determine stability and shape of the clusters.
The ionic skeleton of alkali metals is soft, 
and cluster geometries are often solely determined by electron correlations.  
In quantum dots and rings, the electrons may be confined
by an external electrostatic potential, formed by a  
gated heterostructure.  In the low density limit, the electrons may 
form so-called Wigner molecules, for which the many-body quantum spectra  
reveal the classical vibration modes.  
High rotational states increase the tendency for the electrons to localize.  
At low angular momenta,  the electrons may form a quantum Hall liquid 
with vortices. In this case, the vortices act as 
quasi-particles with long-range effective interactions that localize 
in a vortex molecule, in much analogy to the electron localization at 
strong rotation.  
\end{abstract} 
 
\maketitle 
 
\section{Introduction} 
 
For a long time, the homogeneous electron gas has been 
the standard theoretical  model for a correlated, 
infinite Coulombic system where the fermionic character 
of the electrons plays the dominant role~\cite{fetter1971}. 
The so-called 'jellium' model of a metal has been a starting point for  
developing functionals for the density functional theory of 
electrons~\cite{kohn1965}.  
The long-ranged nature of the Coulomb interaction has posed a challenge to
many-body theory since many decades. 
At high densities of the electron gas, the exchange interaction 
dominates, while in the low-density limit, the electrons may 
form a Wigner crystal~\cite{wigner1934}. 
 
In real metals, the ions do not form a homogeneous charge background 
like in the simple jellium model, but may rather be described by a lattice  
of pseudo-potentials. Nevertheless, many properties of alkali  
metals can be understood on the basis of the jellium model, 
as for example, the surface energy and work function~\cite{lang1970},  
the vacancy formation energy~\cite{manninen1975},  
or collective plasmon excitations~\cite{fetter1971}, etc. 
Since this works well for the bulk, it is not 
surprising that the jellium model also applies well to the approximate
description of the properties of small alkali metal 
clusters~\cite{heer1993,brack1993}. 

\smallskip

The intention of this article is to provide a brief 
survey of some of the fascinating properties of clusters or quantum
dots: 
In Section 2, we discuss  
metal clusters on the basis of a two-component plasma and show that 
the overall shape and the plasmon excitations can be  
qualitatively explained within this simple model. 
In semiconductor heterostructures, the valence electrons may be confined 
within a quasi two-dimensional layer, 
forming two-dimensional electron gas (2DEG).  
Here, depending on the material parameters, the electron density 
is small, and the electron wave length much larger 
than the lattice constant. 
Consequently, the model of the two-dimensional 
homogeneous electron gas is valid, if the electron mass is 
replaced by an effective mass determined by the band structure, 
and the Coulomb interaction is replaced by an interaction screened by 
the dielectric constant of the semiconductor~\cite{reimann2002}.  
With etching techniques and external gates, electrons can 
be localized in a nearly harmonic confinement.  
In Section 3 we will study the fascinating many-particle 
physics of this seemingly simple system: 
one example of its surprisingly rich physics 
is the occurrence of internally broken symmetries such as 
spin density waves in the ground state. 
The following sections concentrate on the localization of electrons 
confined in quantum rings or 2d quantum dots. We will see how the 
reduction of the dimensionality increases 
the electron-electron correlation, and how the Pauli exclusion principle 
then comes to play the dominating role. 
Using a simple model for a quantum ring, we show in Section 4 
that the many-particle 
energy spectrum reveals the internal structure of the ground state, 
and its analysis provides information about the Wigner localization.  
The collective excitations of the particles  
are then the classical vibrational modes of the Wigner molecule. 
Correspondingly, in section 5 we then show how all the low-energy 
states at high angular momenta can be described  
by a simple model Hamiltonian of a vibrating molecule.  
In Section 6 we turn to a discussion of collective excitations 
at low angular momenta. For large numbers of electrons, the 
low-energy excitations correspond to vortices in the quantum system.   
The localization of vortices in a 'vortex 
molecule' then determines the fine structure of the  
energy spectrum. 
We connect the discussion to the physics of  
other, but intimately related many-body systems, such as for example 
cold atoms in traps, pointing out the apparent similarities between 
finite fermionic or bosonic quantum systems.

\section{Metal clusters as electron-ion plasma} 
 
Mie~\cite{mie1008} showed already 100 years ago that a metal sphere has 
an optical absorption at a frequency which is independent of the  
size of the sphere. The absorption peak corresponds to a  
surface plasmon related to the bulk plasmon 
as $\omega_{\rm sp}=\omega_{\rm p}/\sqrt{3}$. 
Since the simplest model for a metal cluster is a conducting sphere,  
it is apparent that similar plasmon peaks as those predicted by Mie, 
were observed~\cite{brechignac1992}. 
As mentioned above, the jellium model assumes the ions to be distributed 
in a homogenous, rigid charge background. Modeling the cluster as a 
jellium sphere, the conduction electrons are then confided in the cluster by 
the Coulomb attraction of the 
sphere~\cite{martins1981,hintermann1983,ekardt1984}.  
The confining potential inside 
the sphere is harmonic, meaning that the center-of-mass motion of the 
electron cloud can be separated from the internal motion. The  
collective oscillation of the electrons against the background charge 
is the plasmon of the metal sphere. This result, in fact, emerged 
also from a shell-model type of calculation for the correlated 
electrons in the cluster~\cite{koskinen1994}. 
The phenomenon is analogous to the giant resonance in nuclei,  
where the protons oscillate against the neutrons~\cite{bohr1975}. 
A more detailed calculation shows that the electrons spill out  
slightly from the region of the harmonic well. 
This causes a small red-shift of the plasmon peak   
(for a review, see~\cite{brack1993}).  
Experiments with large spherical alkali-metal clusters 
have shown that the plasmon peak occurs in fair agreement with this  
simple model~\cite{brechignac1992}. 
 
The spherical jellium model predicted a shell structure for small  
sodium clusters which was observed first by 
Knight {\it et al.}~\cite{knight1984}. 
Immediately after this discovery, however, it was realized that 
only metal clusters with filled electron shell are nearly spherical, 
while others should have strong  
(quadrupole) deformations, in much analogy to the physics of 
atomic nuclei~\cite{clemenger1985}. 
Below, we shall discuss the origin of this shape deformation by 
considering the sodium metal as a two-component plasma consisting 
of ions and electrons~\cite{manninen1986}. The ions are mimicked by 
positrons (but shall not allow annihilation with electrons).  
A cluster of such a fictive 'electron-positron' 
plasma can be studied using the density functional Kohn-Sham  
method, with the single particle equations 
\begin{equation} 
-\frac{\hbar^2}{2m} \nabla^2\psi_i + v_{\rm eff}\psi_i=\epsilon\psi_i, 
\end{equation} 
where the effective potential is the functional derivative of the  
potential energy functional $V[n]$: $v_{\rm eff}({\bf r})=\delta V[n]/\delta n({\bf r})$. 
In the case of our imaginary electron-positron system, the potential energy 
consists only of the exchange and correlation energies since the total  
Coulomb potential (Hartree term) is zero due to the symmetry 
(the electron and the positron densities are identical).  
Alternatively, we can imagine the positive ions forming a completely 
deformable, 'floppy' (classical) background charge which will always  
take the same density distribution as the electron density, but does 
not have its own correlation or exchange energy. This is the  
so-called 'ultimate jellium' model~\cite{koskinen1995} where the 
potential energy is now {\it only} the exchange and correlation energy 
of the electron gas, $V[n]=E_{\rm xc}[n]$. The equilibrium density 
of the infinite ultimate jellium is close to that of the electron density 
of sodium (with a Wigner-Seitz parameter of $r_s\approx 4.2 a_0$, where $a_0$
is the Bohr radius). Nevertheless, it  
is still surprising how well the model can quantitatively describe 
shape-related properties of real sodium clusters~\cite{moseler2003}. 
The obtained shapes are in agreement with those determined by the splitting 
of the plasmon resonance to two separate peaks~\cite{borgreen1993}.  
 
\begin{figure}[h] 
\hfill\includegraphics[width=0.6\columnwidth]{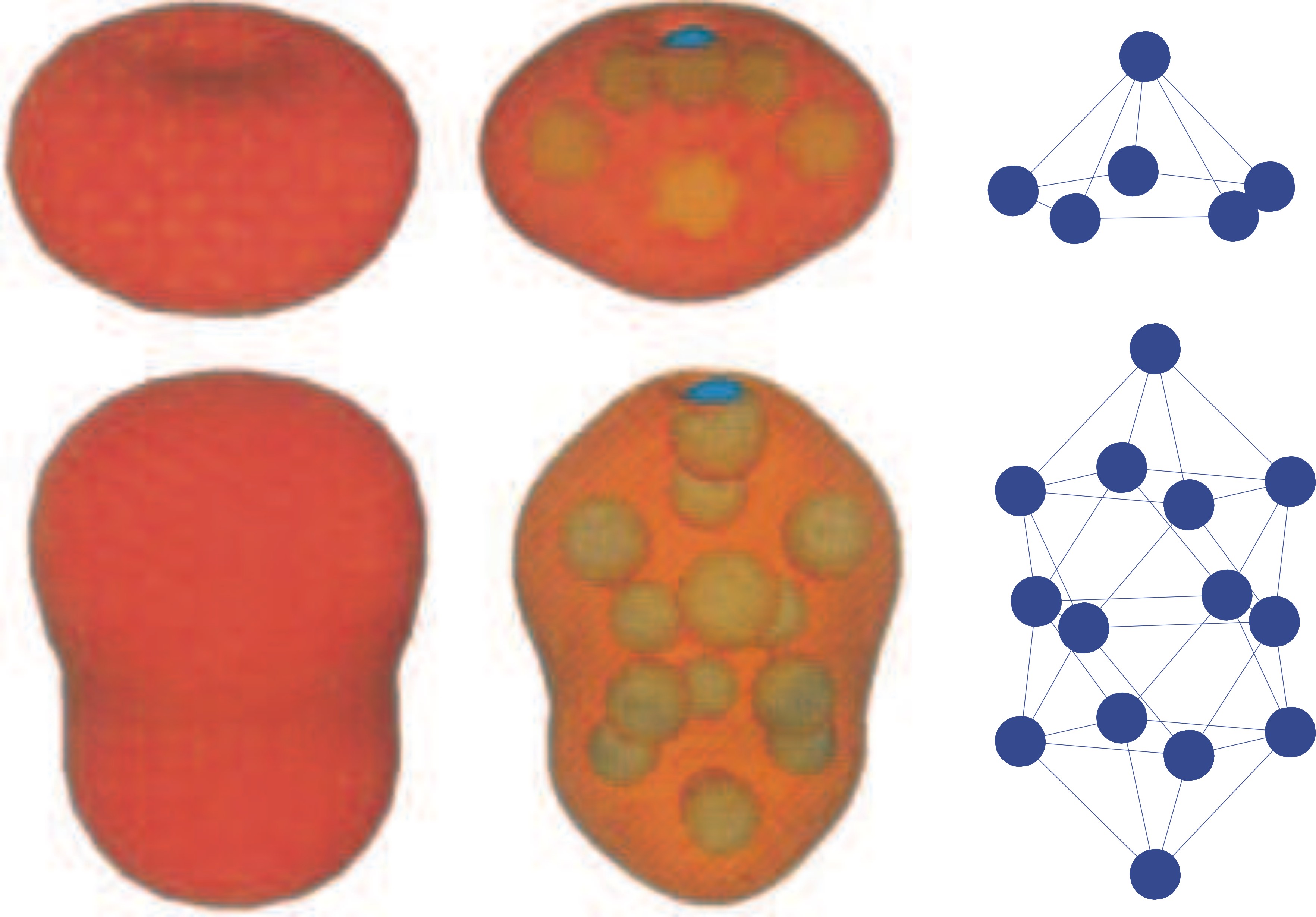} 
\caption{Shapes of clusters with 6 and 14 particles. The left panel shows 
the constant-density surface of the universal model giving similar 
shapes irrespective of the interparticle interactions.  
The center panel shows the constant density surface and ion positions 
in sodium clusters calculated within DFT. The right panel shows  
stick-and-ball models of atom positions of TB clusters. Note that in this 
case the constant density surface can not be defined, but the 
resulting atom positions are the same as in the DFT calculation. 
} 
\label{clusters} 
\end{figure} 
 
The success of the simple plasma model for describing the overall shapes 
of sodium clusters is based, on one hand, on the softness of the metal,  
and on the other hand, on the universality of 
the shapes of small fermion systems~\cite{hakkinen1997}. 
In small systems, there are only a small number of single particle states 
which contribute to the density distribution. The effective potential 
is a functional of this density distribution and 
will thus have a similar shape determined by the few single-particle 
wave functions, irrespective to the details of 
the interparticle interactions binding the system together.  
Figure~\ref{clusters} illustrates the strength of this universality, showing   
the results for three a priori very different models for 
clusters containing 6 and 14 particles.
The density contours of the two clusters on the left are results of  
the universal model (or so-called 'ultimate jellium' model), 
where the only essential parameter is the average density of the bulk 
material. The clusters shapes shown in the center panel are the 
results of ab-initio density functional calculations 
using ab initio pseudo-potentials. The right panel shows the 
results of the simplest possible tight-binding model, which 
assumes constant bond length and only nearest neighbor hopping.  
Note that this tight binding model does not take into account any long-range  
Coulomb interactions.   

\smallskip

In semiconductors, is possible to make two-dimensional 
structures where electrons 
and holes are at different layers, the separation of the layers 
hindering their recombination. 
In the limit of vanishing interlayer distance the electrons and holes 
form a two-dimensional plasma. Reimann {\it et al.}~\cite{reimann1998} 
studied finite-size systems of such a 2D plasma and observed that, 
like in 3D, the  shapes of the plasma clusters are quite robust. 
Interestingly, in 2D those clusters corresponding to filled electronic 
shells of 2D  circular traps do not have circular shapes but  
strong triangular deformations, as shown in Fig. 2. The reason is that 
in a triangular cavity the lowest energy shells agree with those of the  
harmonic oscillator~\cite{reimann1997}.  
 
\begin{figure}[h] 
\hfill\includegraphics[width=0.8\columnwidth]{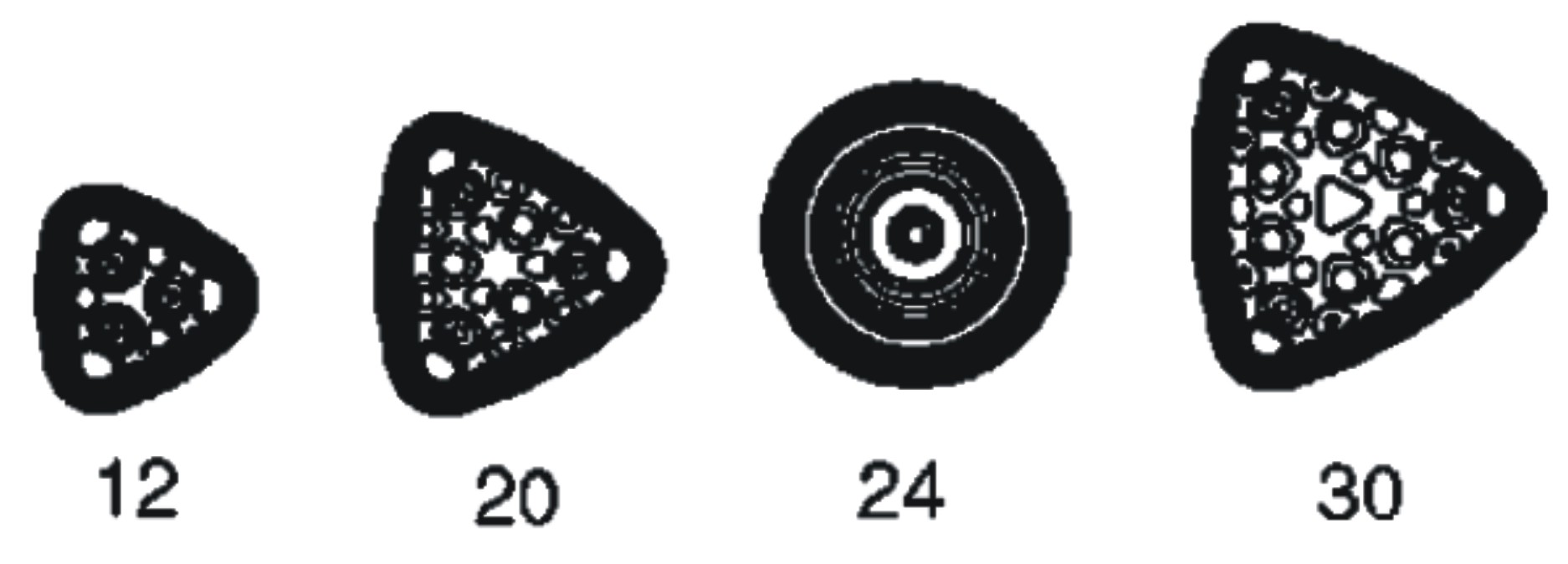} 
\caption{Electrons densities in 2D plasma clusters  
calculated with the ultimate jellium model. Note that the 
sizes 12, 20, and 30 correspond to filled shells. 
} 
\label{tt} 
\end{figure} 
 
It has also been suggested~\cite{kohl1996,kolehmainen1997} that the  
two-dimensional deformable jellium 
model could capture the main correlations of alkali metal 
clusters on a weakly interacting substrate, like oxide or 
graphite. In such systems computations based on DFT and  
pseudo-potentials~\cite{hakkinen1997} indeed also result in a  
triangular shape for $N=12$ in agreement with the simple model. 
 
\section{Two-dimensional quantum dots: DFT} 
 
In the previous sections, we studied free metal and plasma clusters  
and learned that, just like in nuclei, any open-shell system 
is deformed with respect to the spherical shape. In 2D, even closed-shell 
clusters do not necessarily occur only for circular shape. 
The internal symmetry breaking 
is driven by the tendency to maximize the energy gap between the 
highest occupied and lowest unoccupied single particle state, 
in accordance with the Jahn-Teller theorem. 
In the remaining parts of this brief survey, let us consider 
systems where the electrons (or ions and atoms, respectively) 
are trapped by a rigid harmonic confinement. 
Physical examples of such systems are the conduction 
electrons in semiconductor quantum dots, as well as atoms and ions 
confined in magneto-optical traps.  
 
Generally speaking, with interactions between the particles, 
the degeneracy of an open shell can be reduced  
by spin polarization driven by Hund's first rule. 
However, this mechanism may be competing with other, internal, 
symmetry-breaking. Examples are deformation effects (as  
discussed above for metal clusters and the jellium model), 
pairing (like in nuclei for interactions with an effectively attractive part), 
or the formation of a spin density wave (see below).
 
Tunneling spectroscopy of small quantum dots~\cite{tarucha1996} 
has clearly revealed the energy-lowering of the half-filled shells  
due to Hund's first rule. The results are strongly supported by  
the spin-density functional~\cite{koskinen1997} and ab-initio many-particle 
calculations for electrons in a harmonic confinement (for a review 
see \cite{reimann2002}). 
 
In the low-density electron gas, the electrons localize in a Wigner  
crystal. Close to this limit, the difference of the total energy 
of the paramagnetic and ferromagnetic electron gas diminishes. 
When the electrons localize, the spin ordering  
is expected to follow the Heisenberg model for the spin. 
Density functional theory in the local density approximation 
can not describe properly the Wigner crystal, since the Coulomb  
self-interaction of the localized electrons is not properly  
canceled by the local exchange. In order to study particle 
localization, despite of an a-priori lack of correlation, it was argued 
that it is then more favorable to use
the unrestricted  Hartree-Fock method~\cite{yannouleas1999}. 
 
\begin{figure}[h] 
\hfill\includegraphics[width=0.6\columnwidth]{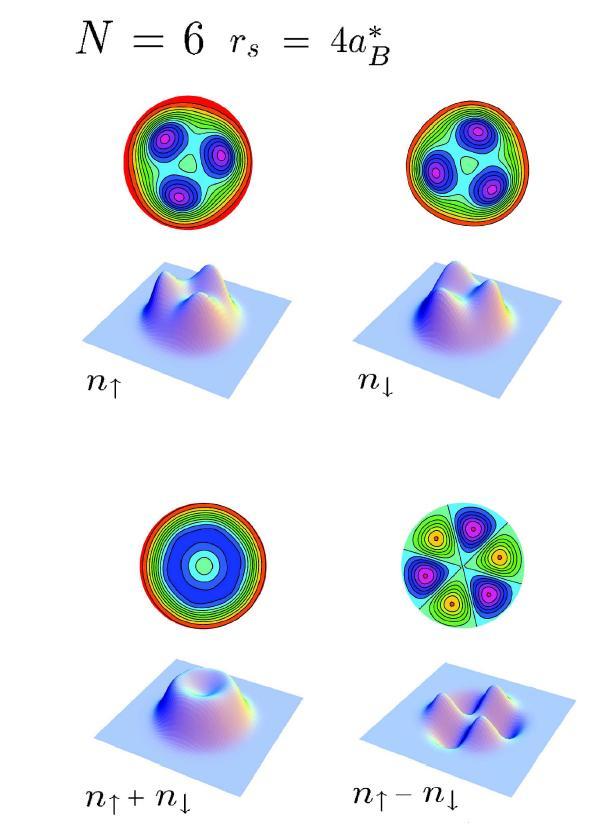} 
\caption{DFT spin densities $n^{\uparrow }$ and $n^{\downarrow }$ 
(upper panel) and total density ($n^{\uparrow }+ n^{\downarrow }$) 
as well as (un-normalized) spin polarization ($n^{\uparrow }+ n^{\downarrow}$)  
(lower panel) for a six-electron quantum dot at $r_s = 4 a_0^*$,   
shown as 3D plots and their contours.  
From Ref.~\cite{borgh2005}.
} 
\label{spinwave} 
\end{figure} 

At large densities, i.e. $r_s \le 6 a_0^* $ (here 
$a_0^*$ is the effective Bohr radius), the six-electron quantum dot has a
closed-shell configuration. Its ground state is a state with total spin
$S=S_z=0$ in both the CI and SDFT methods, with circularly symmetric 
particle densities. As $r_s$ increases, however, the total density remains 
azimuthally symmetric, while the spin densities show a symmetry breaking,
leading to a pronounced spatial oscillation in the spin polarization
(see Fig.~\ref{spinwave}), but still total spin $S_z=0$.
Such so-called 'spin density wave' (SDW)-like states~\cite{koskinen1997} 
have been much discussed in the literature, and it was claimed that such
states are simple artefacts of the broken spin symmetries in 
SDFT~\cite{hirose1999,harju2004}. To resolve this question, clearly one has to 
compare the SDFT results with the solutions of the full many-body
Hamiltonian.
The many-particle state of a few electrons in a quantum dot can 
be obtained numerically nearly exactly, by diagonalizing  
the Hamiltonian in a properly restricted Hilbert space.  However, 
density and spin densities of the exact solution necessarily have 
the same symmetry as the Hamiltonian. For a circular quantum dot, this means 
that the solutions also must have the azimuthal symmetry of the Hamiltonian.  
Consequently, the possible localization of  
electrons, or other internal symmetry breaking such as the above mentioned 
spin-density waves, can {\it not} be seen  directly in the total particle- or 
spin-densities. Instead, one then has to investigate 
the pair correlation functions, i.e., conditional probabilities, 
to study the internal structure~\cite{reimann2000,yannouleas2004}. 
Indeed, the direct comparison to spin-dependent pair 
correlation functions in the configuration
interaction method, clearly demonstrated that SDW-like states in
question can appear in the internal structure of the exact many-body 
state~\cite{borgh2005}. However, the SDFT results remain questionable 
for the discussion of spin multiplets, as already shown by von Barth in
1979~\cite{vonBarth1979}.  

\smallskip

In a multi-component system, the particles may have also other 
internal degrees of freedom than the spin.  
These can be, for example, electrons 
in different layers of a semiconductor heterostructure,  
or electrons corresponding to different conduction electron minima  
in a multi-valley semiconductor. For many internal degrees of  
freedom, the DFT-LDA can capture even the localization of  
particles in the low-density limit~\cite{karkkainen2004}. 
 
In the frequently applied Monte Carlo method (see~\cite{harju2005} 
for a review) the electron localization can be 
mapped by following the electron paths a finite number of  
Monte Carlo steps~\cite{filinov2001}.  
In the following, we will show that 
the many-particle energy spectrum can be directly used as 
a signature of particle localization. We will first consider the 
simple case of a one-dimensional quantum ring. 
 
\section{Energy spectra and localization: Quantum rings} 
 
In a strictly one-dimensional ring, the single-particle 
energy levels are solutions to the angular momentum-part of the Hamiltonian, 
\begin{equation} 
H=-\frac{\hbar^2}{2mR^2}\frac{\partial^2}{\partial \phi^2}, 
\end{equation} 
where $R$ is the radius of the ring. The solutions are  
$\psi_\ell (\phi)=\exp(i\ell\phi)$ with energy eigenvalues 
$\epsilon_\ell=\hbar^2\ell^2/2mR^2$. For simplicity we will  
consider non-interacting electrons with the same spin. Then 
each electron occupies a different single-particle state, and the 
energy is the sum of the energies of the single-particle energies. 
Figure 4 shows as an example the resulting (here non-interacting) 
many-particle energy spectrum 
for four (spinless) electrons.   
 
\begin{figure}[h] 
\hfill\includegraphics[width=0.8\columnwidth]{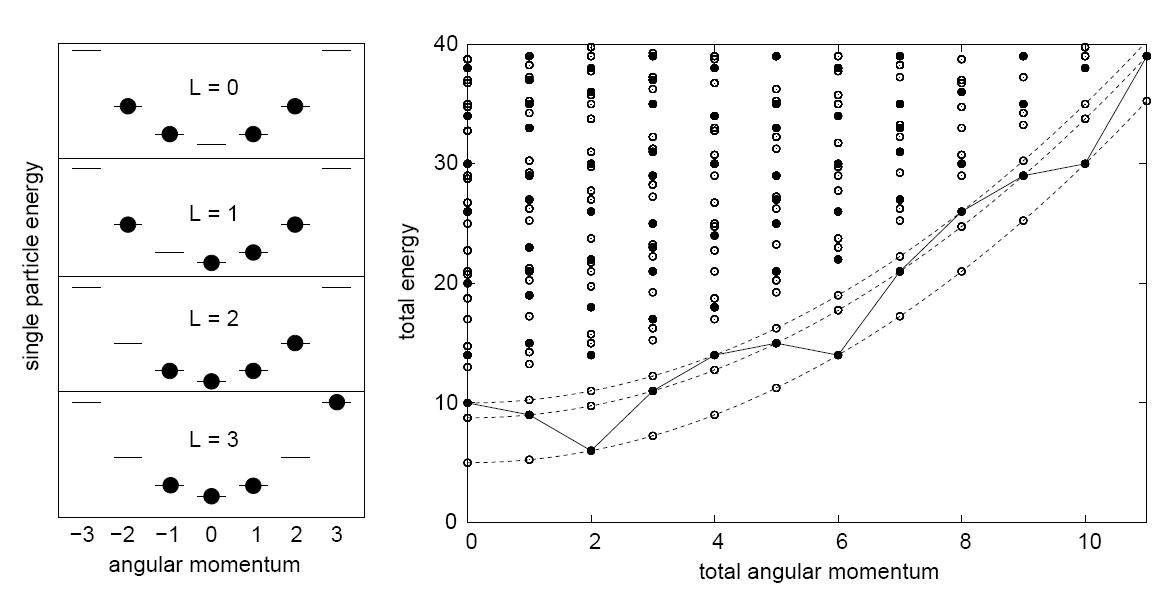} 
\caption{Spectra of one-dimensional quantum rings. 
The right panel shows the spectrum of four 
particles interacting with delta function interaction 
in a strictly 1D ring. Black bullets show results of the  
states with maximum spin ($S=2$), circles have lower spin.  
The solid line connects the lowest states of polarized electrons. 
The dashed lines show the (lowest) rigidly rotating state, and 
the lowest vibrational states. 
The left panel shows the configurations of lowest-energy 
states of polarized electrons for $L=0\cdots 3$, 
demonstrating how the lowest energy state is obtained for $L=2$. 
} 
\label{nonint} 
\end{figure} 
 
More generally, we can consider particles interacting with an infinitely 
strong contact interaction ($v({\bf r-r'})=v_o\delta({\bf r-r'})$, 
where $v_0\rightarrow\infty$). Note that for polarized electrons, the 
contact interaction then does not play any role since the Pauli 
exclusion principle forbids the electrons to be in the same point. 
The many-particle problem with spin can for contact interactions be  
solved exactly with the Bethe ansatz~\cite{lieb1968,viefers2004}. 

Figure~\ref{nonint} shows that for non-polarized
electrons, the lowest energy of each 
angular momentum (the so-called yrast state) 
is a smooth function of $L$, while for polarized electrons it 
oscillates with a period of four. The reason for this  
oscillation becomes obvious when one considers 
the configurations of each of these states, as  
illustrated in the figure. If there is a non-occupied 
state between the occupied ones, the energy is higher than for  
the compact states. The period of four is a result of the following fact: 
If $\Psi_L$ is a solution for a ring of $N$ particles with angular momentum 
$L$, then also 
\begin{equation} 
\Psi_{L+\nu N}=\exp\left(i\nu\sum_k^N \phi_k \right)\Psi_L, 
\end{equation} 
is a solution with angular momentum $L+\nu N$, that also has exactly the same 
internal structure.
One can interpret the eigenenergies in Fig.~\ref{nonint}  
as a rotation-vibration spectrum of 
localized electrons. The states at the minima (lowest dashed line)  
correspond to rigid rotations of localized electrons  
while the states above have vibrational states accompanying the rigid rotation 
(two lowest vibrational states are shown as dashed lines). 
The possibility of vibrational states of noninteracting electrons is  
a peculiarity of the 1D system: An electron is localized between the 
two neighboring electrons since the Pauli exclusion principle prevents 
them to pass each other. This gives an effective $1/r^2$ interaction between 
the electrons which in this case arises from the kinetic energy of the electrons. 
Indeed, quantizing the vibrational models of a ring formed of particles  
with $1/r^2$ interactions gives exactly the spectrum shown in Fig. 4. 
 
In quasi-one-dimensional rings with electrons interacting by  
long-range Coulomb interactions, the  
localization may also be seen directly in the energy spectrum. 
If the electrons are non-polarized, 
the charge and spin degrees of freedom separate. The charge excitations 
are the rigid rotations and vibrations, while the spin excitations  
can be described by the anti-ferromagnetic Heisenberg model of  
localized spins~\cite{koskinen2001} (for a review see~\cite{viefers2004}). 
The conclusion is that in narrow quantum rings the electrons localize 
in a 'necklace' of electrons, and the collective low-energy excitations are 
vibrations and spin excitations. 
 
\section{Rotation-vibration spectrum of electrons in quantum dots} 
 
\begin{figure}[h] 
\hfill\includegraphics[width=0.6\columnwidth]{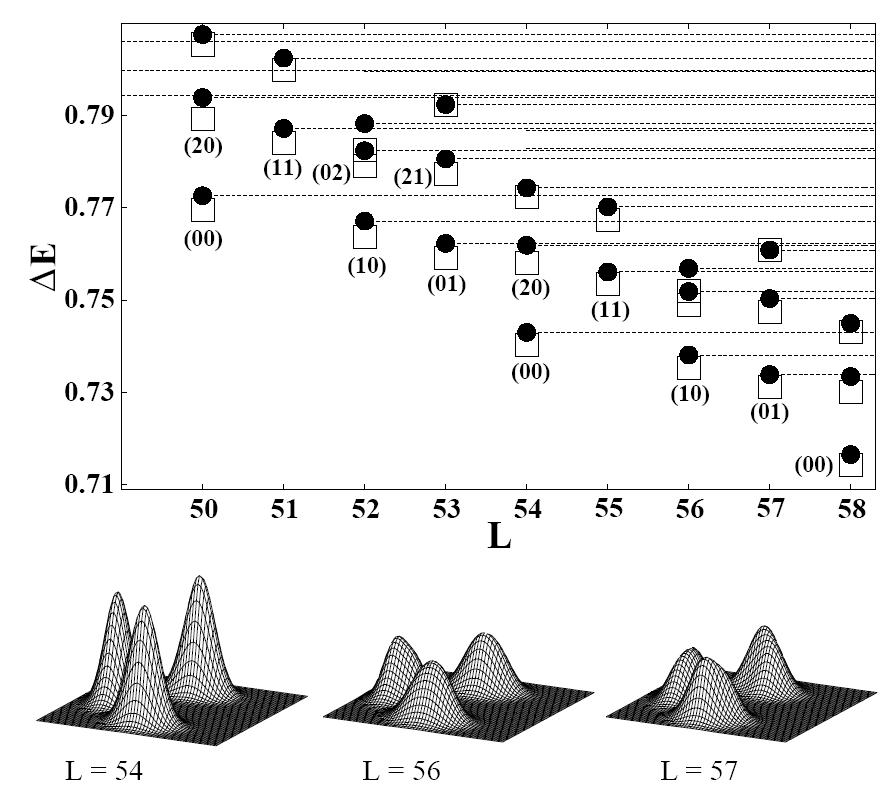} 
\caption{Energy spectrum of four polarized electrons in a 2D quantum dot. 
The black bullets show the result of an exact quantum-mechanical   
diagonalization calculation, while the circles are results of the 
quantization of the vibrations and rotations of classical electrons 
of a Wigner molecule. The numbers ($nm$) show the occupancies of the  
two vibrational modes of the system. The lower panel shows the  
pair-correlation functions of lowest-energy states for three angular momenta. 
The energy is in atomic units and $\Delta E=E_i-L\hbar\omega_0$, where 
$\omega_0=1$ is the confinement frequency. 
} 
\label{pekka} 
\end{figure}

Let us now return to electrons confined in a 2D harmonic potential, and 
consider first the rotational states of polarized electrons with large 
angular momentum. The electrons interact by their long-range Coulomb 
interaction. 
For not too large numbers of electrons, 
the Hamiltonian can be solved numerically 
almost exactly. Maksym~\cite{maksym1996,maksym2000} showed 
that the resulting energy  
spectrum can be quantitatively described by quantizing the classical 
vibrational modes solved in a rotating frame.  
Figure~\ref{pekka} shows that for high angular momenta the whole 
energy spectrum can be quantitatively described by the vibration modes of the  
localized particles.  
 
Figure~\ref{pekka} shows that the lowest-energy state as a function 
of the angular momentum has a similar periodicity of four as observed 
above for four electrons in a ring. 
The figure also shows pair correlation functions 
for three cases. For $L=54$, the pair correlation function 
shows that the electrons are clearly localized: 
Fixing the position of one reference electron fixes the positions of  
the three other electrons. In the cases $L=56$ and $L=57$, the localization 
does not seem to be as strong due to the fact that these states  
correspond to vibrational excitations (for a more detailed analysis,  
see~\cite{nikkarila2007}). 
 
Let us consider the effects of the electron spin on the  
many-particle spectrum and on the electron localization. 
Note that the Hamiltonian of the system does not depend on 
spin (since there is no magnetic field or spin-orbit interaction). 
However, as already noted in connection with the discussion of 
particles on a ring,  the spin-degree of freedom has an important 
role in making the total wave function antisymmetric. 
Moreover, it turns out that in the case of a long-range Coulomb interaction, 
the localized electrons interact with an effective exchange 
interaction as in the Heisenberg anti-ferromagnet.  
In the case of a one-dimensional ring, this can be understood 
on the basis of the half-filled Hubbard model with nearest-neighbor 
hopping\cite{kolomeisky1996,viefers2004}. 
The effective Hamiltonian for electrons localized in a ring can 
be written as 
\begin{equation} 
H_{eff} = \frac{\hbar^2}{2I} {\bf M}^2  
        + \sum_{\nu} \hbar \omega_{\nu} n_{\nu} 
	+ J\sum_{\langle i,j\rangle} {\bf S}_i\cdot{\bf S}_j 
\label{modham} 
\end{equation} 
where $I$ is the moment of inertia, 
$\omega_\nu$ is the eigenfrequency of the 
vibrational mode $\nu$, $J$ is an effective exchange interaction, 
and $S_i$ is the spin operator. 
The value of $J$ becomes smaller when the localization gets more pronounced. 
In the case of the infinitely strong contact interaction 
discussed in the previous section, $J=0$ and the spin/excitations 
have zero energy. For Coulomb interactions $J$ is always finite, 
leading to finite spin excitations. 
 
\begin{figure}[h] 
\hfill\includegraphics[width=0.5\columnwidth]{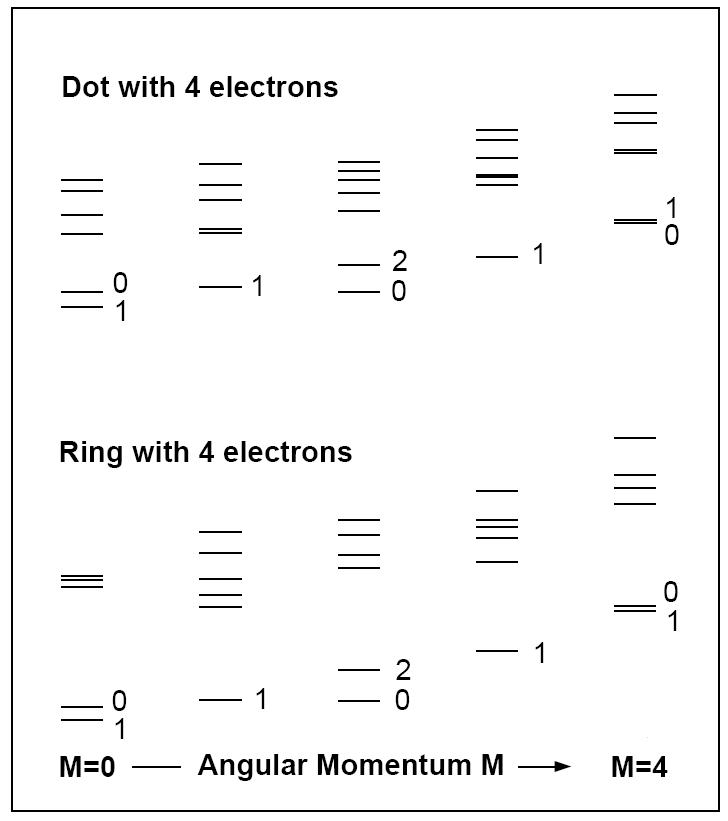} 
\caption{Many-particle energy spectrum for four electrons 
in a 2D quantum dot and in a quantum ring.  
The numbers next to the lowest-energy levels show the 
total spin of the state. 
} 
\label{ringdot} 
\end{figure} 
 
Figure~\ref{ringdot} illustrates the effect of the spin  
in the cases of a 2D dot and a quasi-1D ring with four electrons. 
In both cases, the lowest band corresponds to a rigid rotation 
of the electrons localized in a square. An energy gap separates these 
states from the vibrational excitations. 
In both cases, the low-energy spectrum is similar and consistent 
with the model Hamiltonian. The antisymmetry requirement 
dictates which spin-state is allowed at the given
angular momentum. Group theory can be used to resolve the allowed states. 
We note that the low-energy state for the maximum spin ($S=2$)  
appears at angular momentum 2, in accordance with the simple model 
for quantum rings shown in Fig.~\ref{nonint} above.
The energy differences between the different spin states in the  
ring is consistent with the Heisenberg Hamiltonian 
(last term in Eq. (\ref{modham})). For the dot, the  
order of the spin states for angular momentum 4 is different, 
most likely due to the fact that in this case there is also an  
exchange interaction between the opposite corners of the square 
of the four electrons, omitted in the simple model. 
 
\section{Localization of vortices} 
 
In the previous section we showed that the electron localization 
in a quantum dot shows characteristic features in the 
rotational spectrum. Especially, in the case of a few particles the 
geometry of the localized molecule, i.e. the symmetry group, determines  
the periodic features of the spectrum as a function of the angular momentum. 
The same method can be used to study localization of vortices 
in 2D quantum dots. A strong magnetic field will polarize the 2D electron 
gas in the quantum dot and put it in a rotational state.   
At the angular momentum $L=N(N-1)/2$, the rotating electrons 
form a quantum Hall liquid (QHL) with filling factor 1.  
In the case of a quantum dot, this state is usually called  
the 'maximum density droplet' (MDD)~\cite{macdonald1993}. 
If the magnetic field is increased, the angular momentum grows and 
vortices form in the system. 
The vortices effectively have a repulsive long-range interaction between them, 
which causes their arrangement in a regular geometric pattern, such as it 
has been seen experimentally in the case of 
superconductors~\cite{tinkham1995},  
as well as dilute atom gases in traps~\cite{madison2000}. 

\begin{figure}[h] 
\hfill\includegraphics[width=0.8\columnwidth]{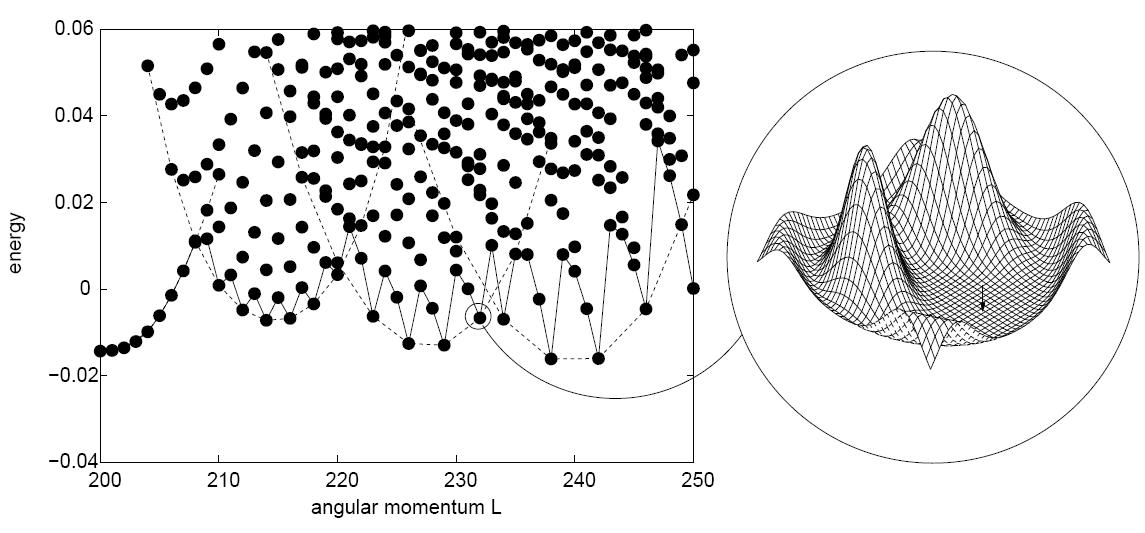} 
\caption{Energy spectrum of 20 polarized electrons 
in 2D a quantum dot showing the periodic oscillations 
arising from localization of two, three and four vortices, 
when the angular momentum is increased. The hole-hole 
pair-correlation corresponding one of the states is shown on the right, 
showing the localization of the three vortices (the reference vortex 
is fixed at the point of the arrow). 
} 
\label{vortex} 
\end{figure}

Evidence for the localization of vortices is also found directly 
in the many-particle spectrum of polarized electrons. 
Figure~\ref{vortex} shows the low-energy spectrum of 20 electrons,  
calculated using the exact diagonalization technique. 
A smooth function of $L$ is subtracted from the spectrum to 
emphasize the details of  the yrast line. 
The figure shows that below $L\approx 210$ 
the yrast line is a smooth function. Beyond that value, the yrast 
line oscillates with a period of two, three and four. 
These oscillations result from the existence of 2, 3 or 4 
localized vortices in the system. The lowest points correspond to 
a rigid rotation of the ring of vortices, while the higher  
energies correspond to vibrational excitations of the vortex system. 
 
If the single-particle basis is restricted to the 
lowest Landau level, i.e. to those single-particle states 
of the 2D harmonic oscillator that have no radial oscillations,  
we can use particle-hole duality to further analyze the 
localization of vortices~\cite{manninen2005}.  
In a polarized Fermi system, we can  
describe any quantum state in the occupation number representation 
equivalently by the particles, or the holes. 
In our present system, the holes correspond to the vortices 
and, consequently, the hole-hole correlation function 
describes the spatial correlation between the vortices. 
Figure~\ref{vortex} shows the hole-hole correlation for 
one of the states in the three-vortex region, showing 
clearly the localization of the three vortices at the corners of an
equilateral triangle.  
Let us finally remark that 
the theories of quantum Hall liquids~\cite{laughlin1983,jain1989,jain1998} 
can be used to discover the similarity of vortex formation in  
fermion and boson systems~\cite{borgh2008}.

\section{Conclusions} 
 
This article is {\it not} meant to be a comprehensive review 
of correlation effects in small system of electrons. 
Rather, we have looked at selected case studies where the 
small size and low dimension of the systems emphasizes  
the correlation and the collective motion of the  
electrons. 
 
In the first section, we showed that in small clusters of 
simple metals it is the optimal shape of the electron  
cloud that dictates the overall geometry of the cluster, 
making the ions the electrons' slaves. The electron-ion 
system behaves like a structureless plasma, where 
the electron-electron exchange and correlation energy determines 
the shape of the cluster. In small systems, this correlation  
is so strong that, in fact, it does not matter what kind of  
internal interaction the particles have, or what 
kind of physical model is used for the system, as  
illustrated in Fig. \ref{clusters}. 
 
A strong external confinement hinders the spatial deformation.
At low densities or high rotation,  
the electrons tend to localize in Wigner molecules. 
The energy spectrum is then dominated by the  
rigid rotation and the internal vibrations of the molecule. 
The localization also separates the spin-excitations 
from the vibrational charge excitations, as most clearly  
seen in quasi-1D rings. A model Hamiltonian consisting of 
rigid rotations, quantized vibrations and an anti-ferromagnetic 
Heisenberg model, describes well the low energy spectrum  
of the system. All low-energy excitations are thus collective  
excitations of strongly correlated electrons. 
  
At high magnetic fields the electrons in a quantum dot 
form a 'miniature' quantum Hall liquid of polarized 
electrons. In this case, the elementary collective excitations 
are vortices. Interestingly, the vortices have a similar 
energy spectrum as the localized electrons. For example,   
the spectrum of three localized electrons shows the same  
vibrational modes as the spectrum of three vortices. 
The vortices can be interpreted as holes 
in the occupied Fermi sea, and the hole-hole correlation function 
can be used to confirm the localization of the vortices.

\end{document}